\documentclass[aps,prd,eqsecnum,nofootinbib,twocolumn,floatfix,superscriptaddress]{revtex4}
\usepackage{graphicx}
\usepackage{dcolumn}
\usepackage{times,mathptm}
\newcommand{\beq}{\begin{equation}}
\newcommand{\eeq}{\end{equation}}
\newcommand{\bea}{\begin{eqnarray}}
\newcommand{\eea}{\end{eqnarray}}

\renewcommand{\d}{\delta}
\renewcommand{\l}{\lambda}
\renewcommand{\L}{\Lambda}
\renewcommand{\b}{\beta}
\renewcommand{\a}{\alpha}
\newcommand{\p}{\phi}
\newcommand{\lm}{\l_0}
\renewcommand{\k}{\kappa}

\newcommand{\g}{\gamma}

\newcommand{\m}{\mu}
\renewcommand{\r}{\rho}
\renewcommand{\P}{\Phi}

\newcommand{\s}{\sigma}

\newcommand{\D}{{\Delta}}

\newcommand{\vph}{\varphi}
\newcommand{\oh}{{\textstyle{\frac{1}{2}}}}

\newcommand{\dg}{\dagger}
\newcommand{\non}{\nonumber}

\newcommand{\rf}[1]{(\ref{#1})}
\newcommand{\ra}{\rightarrow}

\newcommand{\FIGURE}[2][v]{\begin{figure}[#1]#2
\end{figure}}
\begin{document}
%
%
\title{Peculiarities in the Spectrum of the Adjoint Scalar
Kinetic Operator in Yang-Mills Theory}

\author{J. Greensite}
\affiliation{Physics and Astronomy Dept., San Francisco State
University, San Francisco, CA~94117, USA}

\author{A.V. Kovalenko}
\affiliation{Institute of Theoretical and Experimental Physics,
B. Cheremushkinskaya 25, Moscow 117259, Russia}
\author{{\v S}. Olejn\'{\i}k}
\affiliation{Institute of Physics, Slovak Academy
of Sciences, SK--845 11 Bratislava, Slovakia}
\author{M.I. Polikarpov}
\author{S.N. Syritsyn}
\affiliation{Institute of Theoretical and Experimental Physics,
B. Cheremushkinskaya 25, Moscow 117259, Russia}

\author{V.I. Zakharov}
\affiliation{Istituto Nazionale di Fisica Nucleare - Sezione di Pisa,
Dipartimento di Fisica ``E. Fermi" - Universita di Pisa, Largo Pontecorvo 3, 56127
Pisa, Italy
}
\affiliation{Max-Planck Instit\"ut f\"ur Physik,
F\"ohringer Ring 6, D-80805 Munich, Germany}
\date{\today}
\begin{abstract}

   We study the spectrum of low-lying eigenmodes of the kinetic
operator for scalar particles, in the color adjoint representation
of Yang-Mills theory.  The kinetic operator is the covariant
Laplacian, plus a constant which serves to renormalize mass.
In the pure gauge theory, our data indicates that the interval between 
the lowest eigenvalue and the mobility edge tends to infinity in the
continuum limit.  On these grounds, it is suggested that the perturbative
expression for the scalar propagator may be misleading even at distance
scales that are small compared to the confinement scale. We also measure
the density of low-lying eigenmodes, and find a possible connection
to multi-critical matrix models of order $m=1$.

\end{abstract}

\maketitle
%
%
\section{Introduction}\label{Introduction}

   In recent years it has been recognized that kinetic operators in
confining gauge theories may have a low-lying spectrum of localized
eigenstates \cite{Maarten,Us,ITEP}.  By ``kinetic operator" we mean, e.g.,
a  Euclidean 
Dirac operator or covariant Laplacian operator in a background gauge field,
with the possible addition of a constant representing a mass term.
The ``mobility edge" is a point in the spectrum between an interval
of spatially localized eigenmodes, and the bulk of eigenmodes
which are extended over the full volume.\footnote{In fact there is
usually an interval of localized states at both the lower \emph{and}
the upper ends of the spectrum, and therefore two mobility edges.}

    In condensed matter physics, it has long been known that the
Hamiltonian of an electron moving in a stochastic potential has a low-lying
spectrum of localized eigenstates \cite{Anderson}.   If the energy at the mobility edge
is higher than that of the Fermi surface, then the material is an insulator \cite{MacK}.
This is because the propagation of a wavepacket is essentially an interference 
effect among energy eigenstates which are spatially extended.  If there are no
available extended modes, then there is no particle propagation
through the medium.
 
    This condensed matter example motivates us to ask the following 
question:  In a confining lattice gauge theory, what is the interval in GeV
(Dirac operator) or $(\mbox{GeV})^2$ (covariant Laplacian) between the
lowest eigenvalue of kinetic operator, and the mobility edge, as we take the 
continuum limit?   If this interval shrinks to zero in physical units, or is populated
by only a finite number of states in the continuum limit, then the relevance
of localized states to particle propagation is questionable.  On the other
hand, suppose the energy interval of localized states goes to infinity, 
in the continuum limit, for a kinetic operator of some type. This fact may then 
have radical implications for gauge theories with matter fields of the corresponding type. 
 
     In this article we investigate the low-lying spectrum of the covariant Laplacian
in the adjoint representation of the SU(2) gauge group.  Up to a constant, the covariant
Laplacian is the kinetic operator for adjoint scalar fields.  The adjoint
representation is chosen because we know, from the work in ref.\ \cite{Us}, that there is already
something odd about the spectrum:  the lowest eigenmode is localized in a region
whose volume goes to zero in physical units (but infinity in lattice units) in the continuum
limit.   The consequences of this fact for the propagation of scalar particles was not so clear.  
In our present paper we extend our study to the full spectrum of localized states.  
Our objective is to (i) find how
the average interval $\D \l_{mob}$ between the lowest eigenvalue and the mobility
edge varies with coupling; and (ii) compute the density of eigenmodes in this interval.  
With those results in hand, we discuss possible consequences for the two-point function
of adjoint scalar fields in confining gauge theories.

\section{Scaling of the Localization Interval}\label{interval}

      The kinetic operator for scalar fields, in the adjoint representation of the SU(2) 
gauge group, is given by
\beq
             K^{ab}_{xy} = -D^{ab}_{xy} + m^{2}_{0} \d^{ab} \d_{xy}
\eeq
where 
\beq
         D^{ab}_{xy} = \sum_{\m} \Bigl[ U^{ab}_\m(x) \d_{y,x+\hat{\m}}
         + U^{\dg ab}_\m(x-\hat{\m}) \d_{y,x-\hat{\m}}  - 2 \d^{ab} \d_{xy} \Bigl]
\eeq
is the covariant lattice Laplacian, with link variables
\beq
         U^{ab}_\m = \oh \mbox{Tr}[\s^a U_\m(x) \s^b U^\dg_\m(x)]
\eeq 
in the adjoint representation,
and $m^{2}_{0}$ may be taken negative.  We compute numerically the low-lying 
eigenvalues and eigenmodes of the covariant Laplacian operator
\beq
               -D^{ab}_{xy} \p^{b}_{n}(y) = \l_{n} \p_{n}^{a}(x)
\eeq
in thermalized lattice configurations, and from these we compute the 
inverse participation ratio (IPR) of each eigenmode
\beq
              IPR_{n} = V \sum_{x} |\p_{n}^{a}(x) \p_{n}^{a}(x)|^{2}
\eeq
From our previous work \cite{Us} we know that there is some range of eigenvalues of the
covariant Laplacian, lying 
between the lowest eigenvalue $\l_{0}$ and the mobility edge $\l_{mob}$, whose
corresponding eigenmodes are localized, and whose IPR's grow linearly
with lattice volume $V$.  For $\l_{n} > \l_{mob}$ the eigenmodes are extended, and
the IPR's are all of $O(1)$ at large volumes.

\FIGURE[htb]{
\centerline{\includegraphics[width=8truecm]{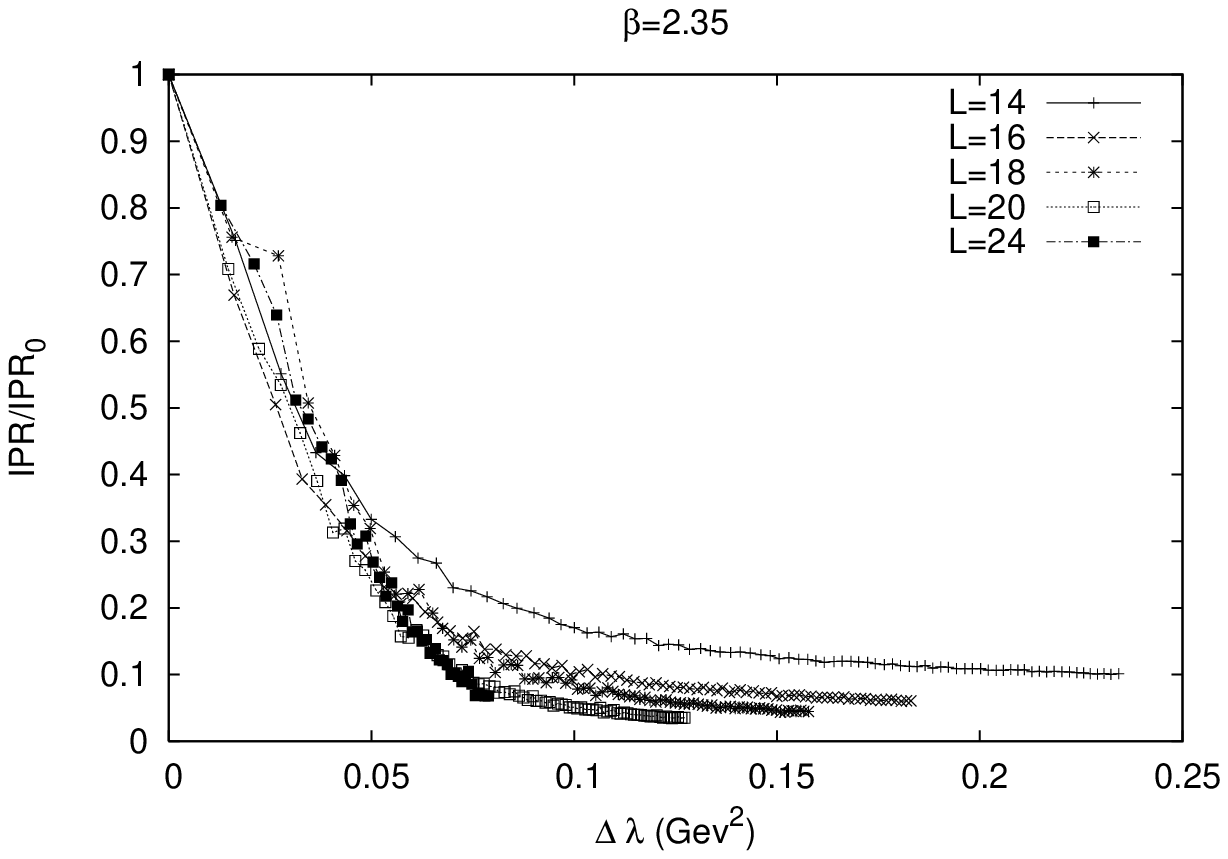}}
\caption{$\langle IPR_{n}\rangle/\langle IPR_{0}\rangle$ vs.\ 
$\D\l_{n} = \langle \l_{n}-\l_{0}\rangle/a^{2}$
at $\b=2.35$, on lattice volumes from $14^{4}$ to $24^{4}$.  The data points
at each volume are connected by lines.} 
\label{b235lines}
}

\FIGURE[htb]{
\centerline{\includegraphics[width=8truecm]{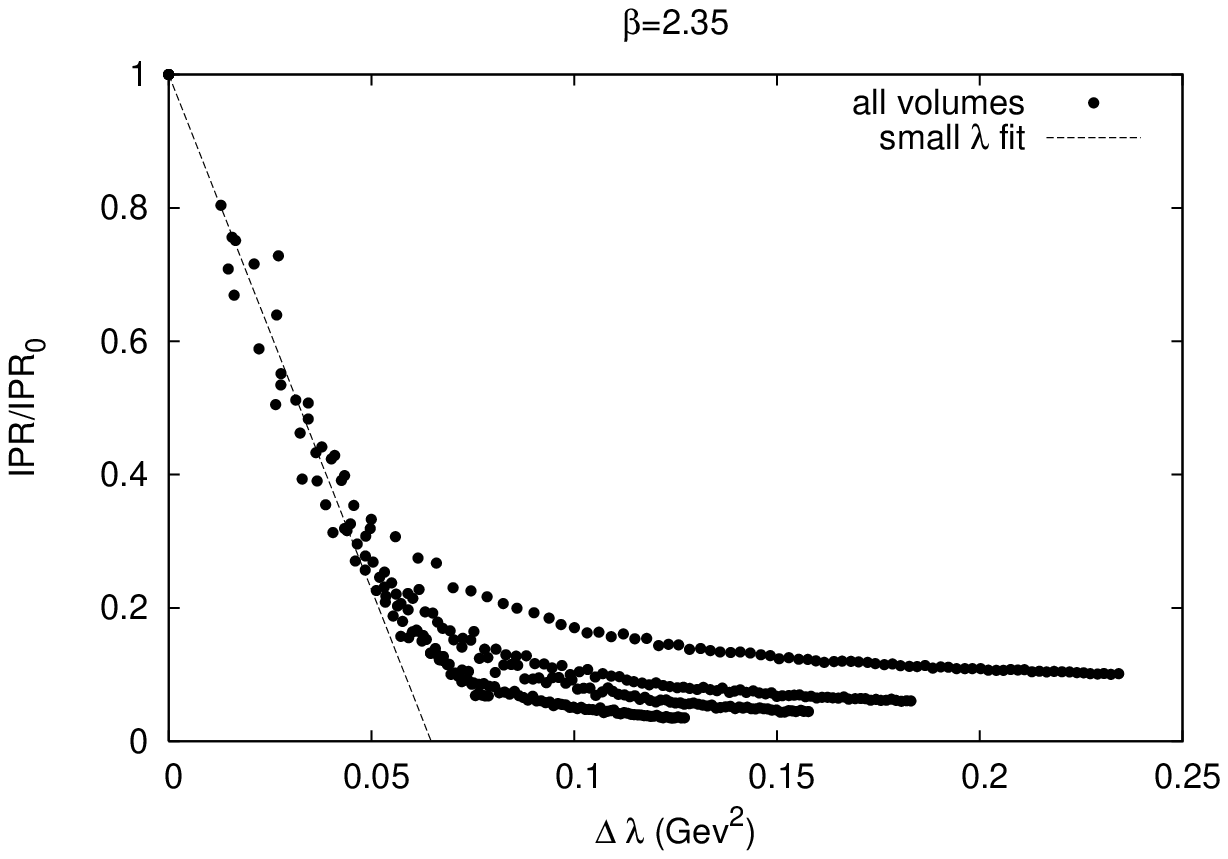}}
\caption{Same as Fig.\ \ref{b235lines}, except lines connecting data
points at different volumes are omitted.  The straight line is a best fit
through the combined all-volumes data in the interval $\D \l \in [0:0.045]$.} 
\label{b235pts}
}
\FIGURE[htb]{
\centerline{\includegraphics[width=8truecm]{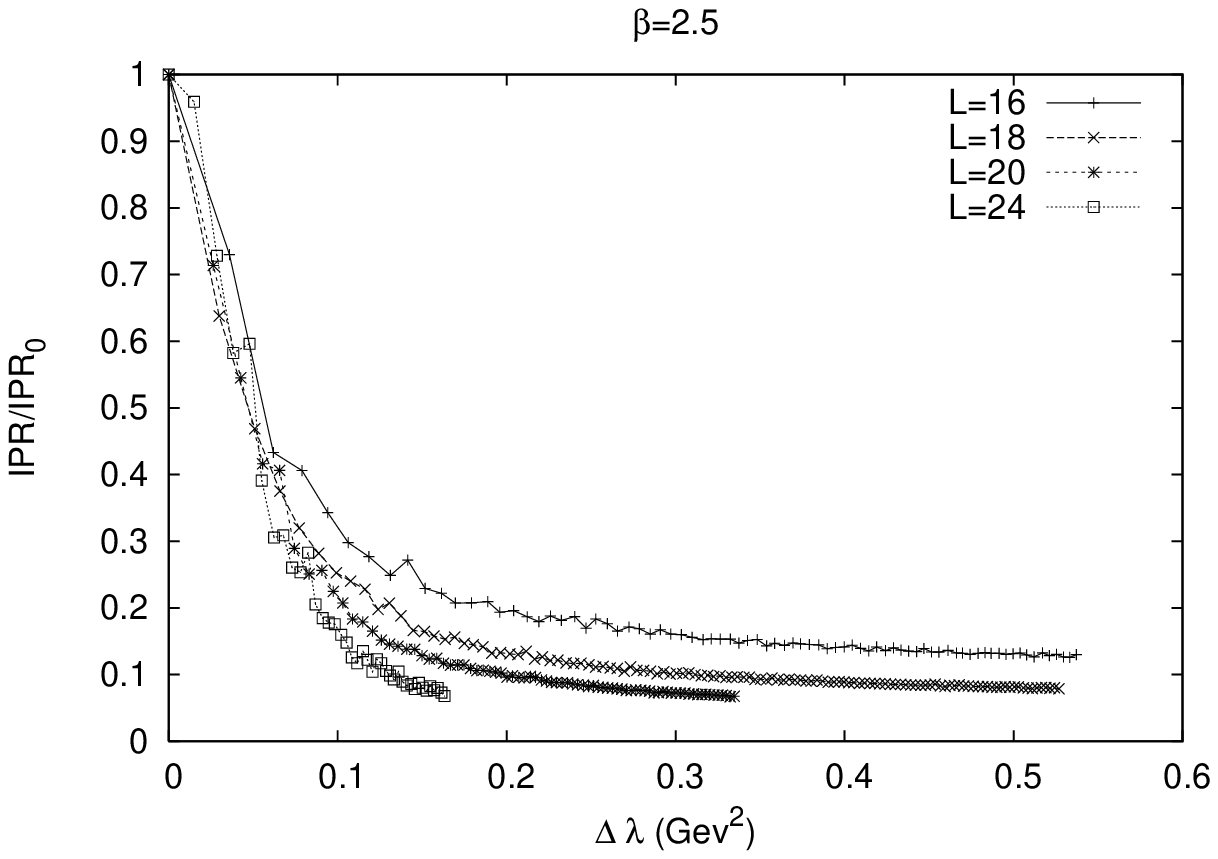}}
\caption{$\langle IPR_{n}\rangle/\langle IPR_{0}\rangle$ vs.\ 
$\D\l_{n} = \langle \l_{n}-\l_{0}\rangle/a^{2}$
at $\b=2.5$, on lattice volumes from $16^{4}$ to $24^{4}$.  The data points
at each volume are connected by lines.} 
\label{b25lines}
}

\FIGURE[htb]{
\centerline{\includegraphics[width=8truecm]{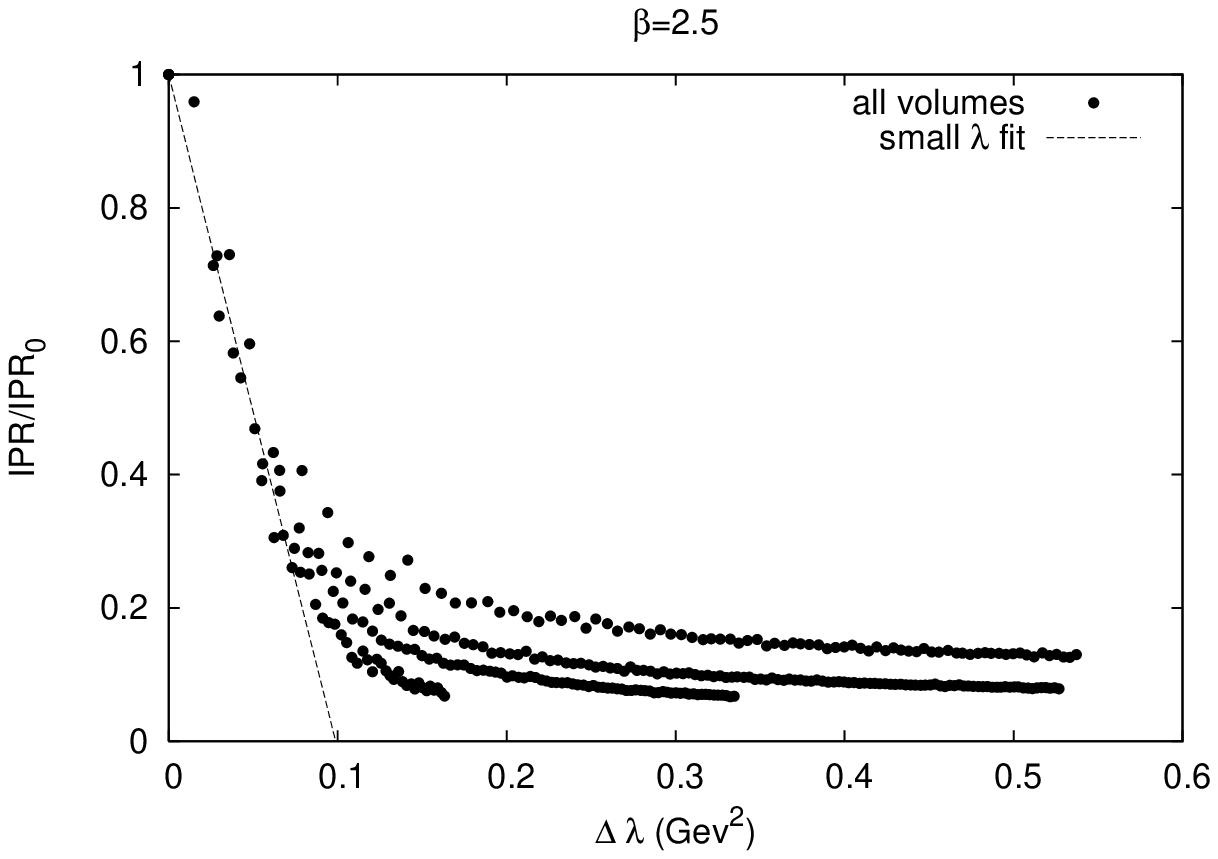}}
\caption{Same as Fig.\ \ref{b25lines}, except lines connecting data
points at different volumes are omitted.  The straight line is a best fit
through the combined all-volumes data in the interval $\D \l \in [0:0.06]$.} 
\label{b25pts}
}

    Let $a$ denote the lattice spacing.  
We are interested in determining how the interval of eigenvalues in physical units
\beq
               \D \l_{mob} = {\langle \l_{mob} - \l_{0} \rangle \over a^{2} }
\eeq
varies as coupling $\b$ increases, and $a\ra 0$.  Our method is based on  
the following observation:  If we plot
\beq
R_{n} = {\langle IPR_{n}\rangle \over \langle IPR_{0}\rangle }
\eeq
vs.
\beq
   \D \l_{n} = {\langle \l_{n} - \l_{0} \rangle \over a^{2}}
\eeq 
at a given coupling $\b$, then the data points at different volumes tend to cluster 
around the same straight line at small $\D \l$, as seen in Figs.\ \ref{b235lines}-\ref{b235pts},
at $\b=2.35$, and Figs.\ \ref{b25lines}-\ref{b25pts} at $\b=2.5$.  
There does not appear to be a strong volume dependence
in the data at $\D \l < \D \l_{mob}$.  Above the mobility edge we would
expect $R_{n} \ra 0$ in the infinite volume limit, since in that limit the IPR's of localized and
extended states tend to infinity and to finite values, respectively.  Fitting a straight line
\beq
f(\D\l)=1- {\D \l \over \D \l_{int}}
\label{fit}
\eeq
to the combined volume data at small $\D \l$, the intercept $\D \l_{int}$ with the $x$-axis is a
reasonable measure of the width of the peak in the $R$ vs.\ $\D \l$ data.  Even if
$\D \l_{int}$ is not precisely the same as $\D \l_{mob}$, we expect these quantities to
scale with $\b$ in the same way.  The result, for $\b\in [2.2,2.5]$ is shown in Table
\ref{table1}, and we see that the data is consistent with
\beq
           \D \l_{int} \approx {0.045(3) \mbox{GeV} \over a }
\eeq
Since $\D \l_{int} \sim \D \l_{mob}$, the implication is that
\beq
            \lim_{a\ra 0} \D \l_{mob} = \infty
\eeq
We reserve discussion of this result to section \ref{discussion}, below.

\begin{table}
\caption{\label{table1} The intercept $\D \l_{int}$ vs.\ $\b$,  as determined
from the best linear fit \rf{fit} to data in the intervals and volumes shown. 
$\D \l_{int}$ and $\D \l_{int} a$ are in units of GeV${}^{2}$ and GeV, respectively.}
\begin{ruledtabular}
\begin{tabular}{|c|c|c|c|c|}
$\b$ & volumes & fitting interval & $\D \l_{int}$ & $\D \l_{int} a$  \\
\hline
2.20   & $12^4,14^4,16^4$           & $[0,0.030]$ & 0.038(1) & 0.042 \\
2.30   & $12^4,14^4,16^4,18^4$      & $[0,0.040]$ & 0.051(1) & 0.044 \\
2.35   & $14^4,16^4,18^4,20^4,24^4$ & $[0,0.045]$ & 0.065(1) & 0.047\\
2.40   & $14^4,16^4,18^4,20^4,24^4$ & $[0,0.050]$ & 0.076(2) & 0.047 \\
2.50   & $16^4,18^4,20^4,24^4$ & $[0,0.060]$ & 0.099(4) & 0.043 \\
\end{tabular}
\end{ruledtabular}
\end{table}
     
\section{Density of Localized States}

       The covariant Laplacian operator in adjoint representation
can  be thought of as a $3V \times 3V$ random matrix,  where the factor of 3 comes 
from the color index.   We are interested in computing the density of localized
eigenstates at the low end of the spectrum.  Suppose this has the form
\beq
             \rho(\l) = \k (\l-\l_{0})^{\a}
\label{rho}
\eeq
where we normalize the density of states such that
\beq
            \int_{\l_{0}}^{\l_{max}} d\l ~ \r(\l) = 3
\eeq
Then the number of eigenmodes with eigenvalues less than $\l$ is
\beq
              n(\l)  = V\int_{\l_{0}}^{\l} d\l'  ~ \r(\l')
\eeq
A standard method for determining $\a$ is based on the observation that under
the eigenvalue rescaling
\beq
                z = V^{1/(1+\a)} (\l-\l_{0})
\label{rescale}
\eeq
the corresponding eigenvalue number $n(z)$ becomes independent of $V$.
This is readily verified:  the number of eigenvalues $\d n(\l)$ in an interval
$\d \l$ around $\l$ is
\beq
              \d n(\l) = V \r(\l) \d \l
\eeq
Rescale $z=V^{p} (\l-\l_{0})$.  Then, using the assumed form \rf{rho} for $\r(\l)$,
we have
\beq
           \d n(z) = \k V^{1-p(1+\a)} z^{\a} \d z
\eeq
Choosing $p=1/(1+\a)$, $\d n(z)$ is volume independent.   So in order to
compute the density of states, we compute $n(\l)$ at a variety of lattice volumes,
and look for a constant $\a$ such that, under the rescaling \rf{rescale}, the curves
for $n(z)$ computed at each lattice volume coincide.

    In numerical simulations, we evaluate the first $N_{ev}$ eigenvalues of the covariant
Laplacian in a set of $N_{conf}$ independent thermalized configurations.
The resulting $N_{conf}\times N_{ev}$ eigenvalues of all $N_{conf}$ configurations are then
sorted from lowest to highest regardless of configuration or eigenmode number.
Then, if $\l_m$ is the $m$-th eigenvalue in the ordered list of eigenvalues, we identify
\beq
           n(\l_m) = {m \over N_{conf}} ~~~~ (m=1,2,...,N_{conf}\times N_{ev})
\eeq
The maximum value is $n(\l_{max})=N_{ev}$, where $\l_{max}$ is the largest eigenvalue in the list.
Other values of $n(\l)$ for $\l<\l_{max}$ and $\l\ne \l_{m}$ are obtained by interpolation (data
points are connected by lines in the figures shown).
The numerical results for $\b=2.5$ on $L^4$ lattices at $L=14,16,20,24$  
are shown in Fig.\ \ref{nlam}.

\FIGURE[htb]{
\centerline{\includegraphics[width=8truecm]{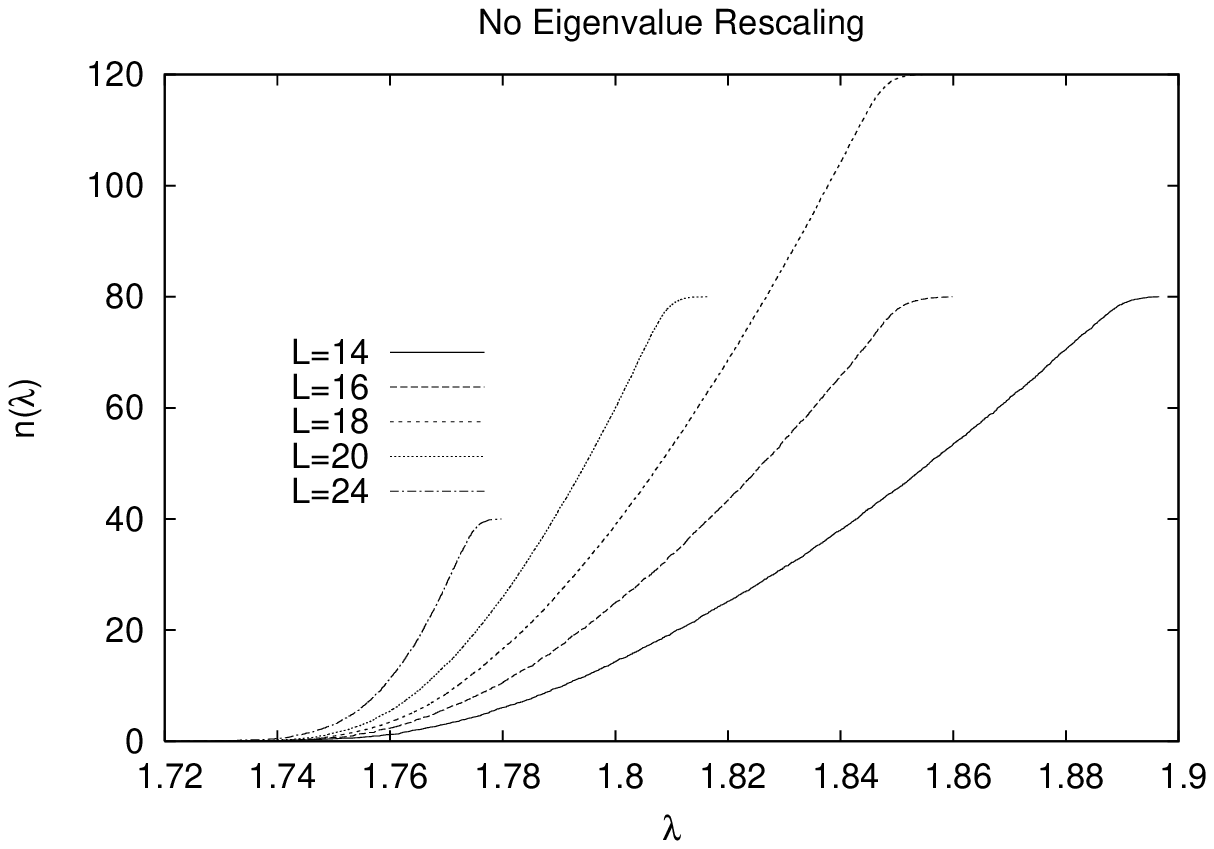}}
\caption{Cumulative number of eigenmodes $n(\l)$ with eigenvalue
less than $\l$ at $\b=2.5$.  Data shown is for lattice volumes from $14^{4}$
to $24^{4}$.} 
\label{nlam}
}

   In Fig.\ \ref{nlam} the $n(\l)$ curve is different for each lattice volume.  This is, of
course, not surprising, since  as volume increases, so does the number of eigenvalues in any given interval
$\D \l$.  Since we are interested in checking universality under rescaling, right at the very end of
the spectrum,  we concentrate on the region where 
$n(\l) < 10$.  In Figs.\ \ref{alpha1}-\ref{alpha4} we show $n(z)$, under the rescaling \rf{rescale}, for
$\a=1,2,4$ respectively.  The data seems compatible with universality at $\a=2$, in which case we
would have
\beq
             n(z) = {\k\over 3} z^{3}
\label{nzfit}
\eeq
Fig.\ \ref{nz} shows a best cubic fit (solid line) to the $\a=2$ scaled data at all volumes,
in the interval $z\in [0,2]$; $\k$ is determined to be $3.8$.  The 
cubic curve appears to fit the data at 
$\b=2.5$ quite well, strengthening the case that $\a=2$ is the correct exponent.

\FIGURE[htb]{
\centerline{\includegraphics[width=8truecm]{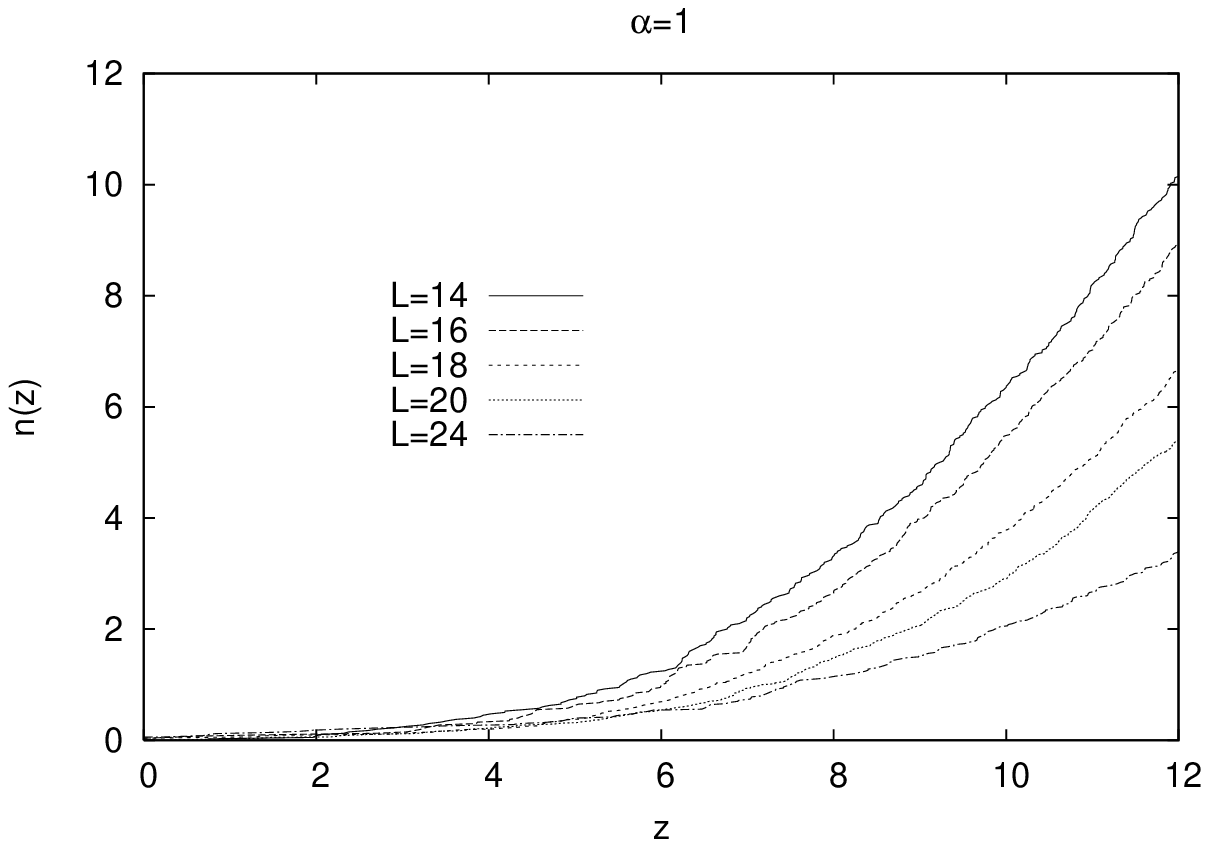}}
\caption{Cumulative eigenvalue number $n(z)$ vs.\ the
scaling variable $z = V^{1/(1+\a)} (\l-\l_{0})$ at $\b=2.5$ and
$\a=1$.} 
\label{alpha1}
}

\FIGURE[htb]{
\centerline{\includegraphics[width=8truecm]{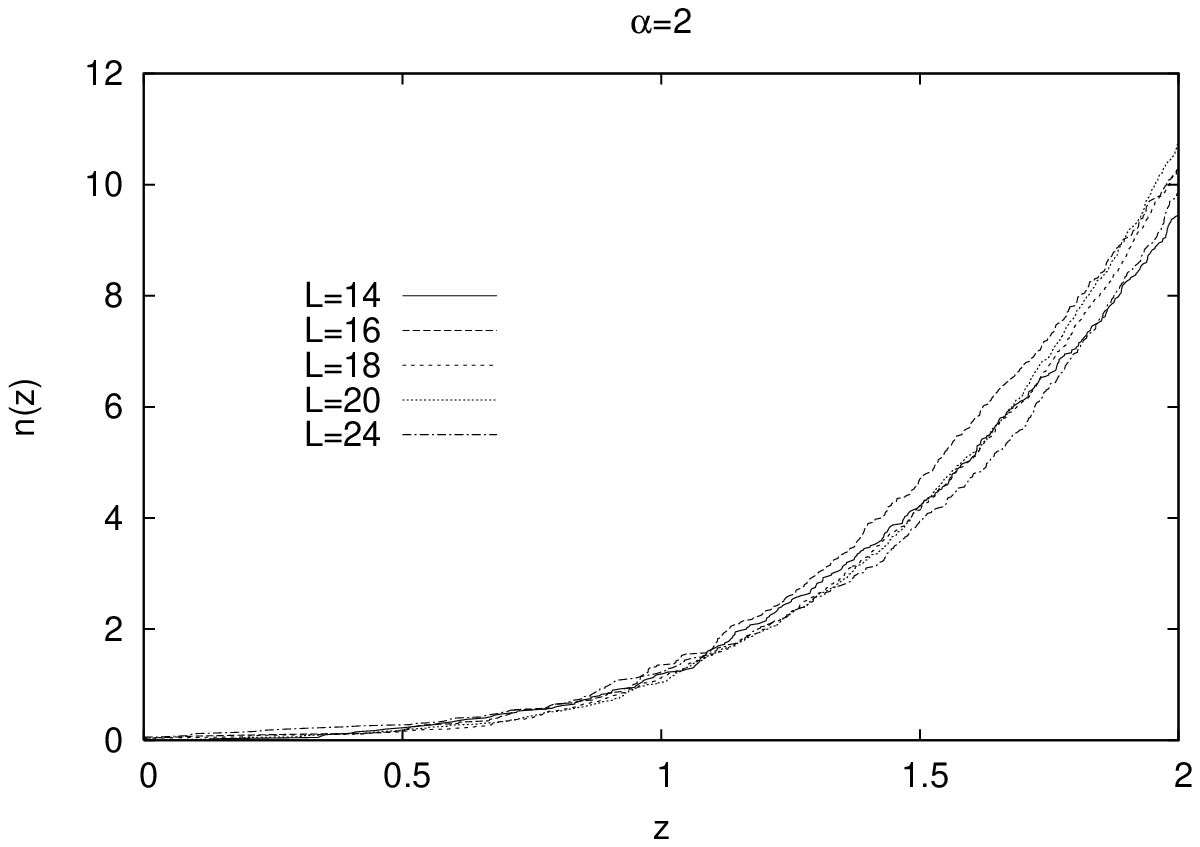}}
\caption{Same as Fig.\ \ref{alpha1}, with $\a=2$.} 
\label{alpha2}
}

\FIGURE[htb]{
\centerline{\includegraphics[width=8truecm]{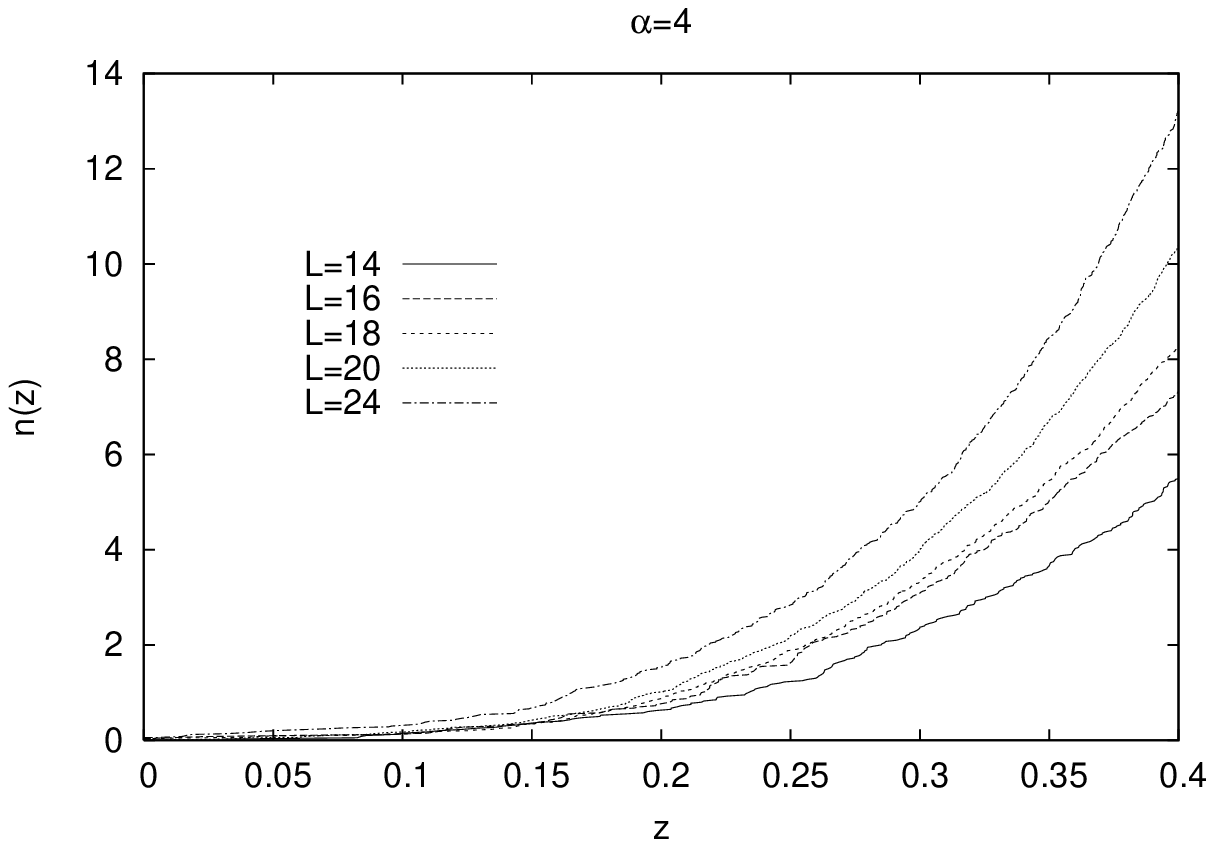}}
\caption{Same as Fig.\ \ref{alpha1}, with $\a=4$.} 
\label{alpha4}
}

\FIGURE[htb]{
\centerline{\includegraphics[width=8truecm]{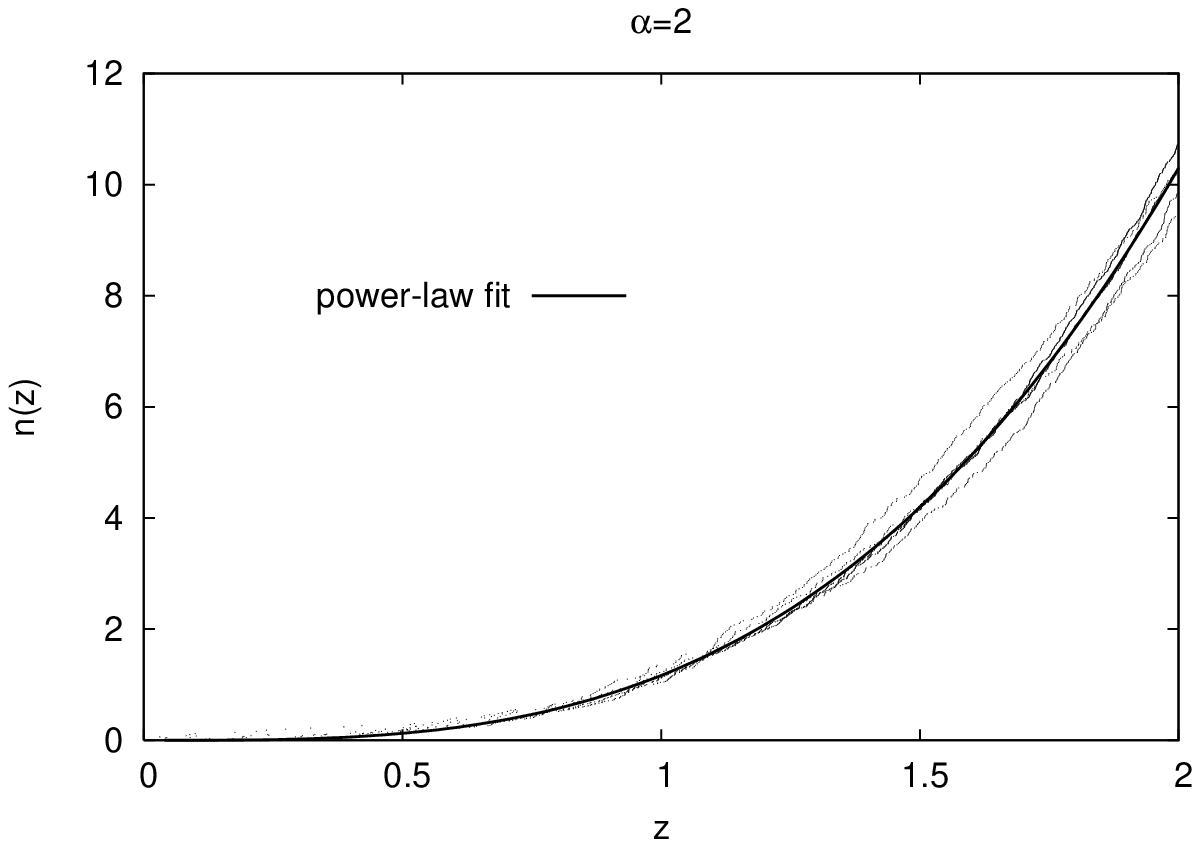}}
\caption{Best fit of $n(z)={1\over 3}\kappa z^{3}$ to the $\b=2.5$ data
(combined volumes) at $\a=2$.} 
\label{nz}
}

    The exponent $\a=2$ may be of some significance.  According to  ref.\ \cite{phd}, 
even-integer exponents $\a=2m$ are obtained in simple
large-N hermitian matrix models with polynomial potentials, when the coupling constants 
in the potential are fine-tuned to achieve criticality.   It appears that the
low-lying spectrum of the covariant Laplacian corresponds to that of a multi-critical matrix model
of order $m=1$.

\begin{figure}[htb]
\centerline{\includegraphics[scale=0.45]{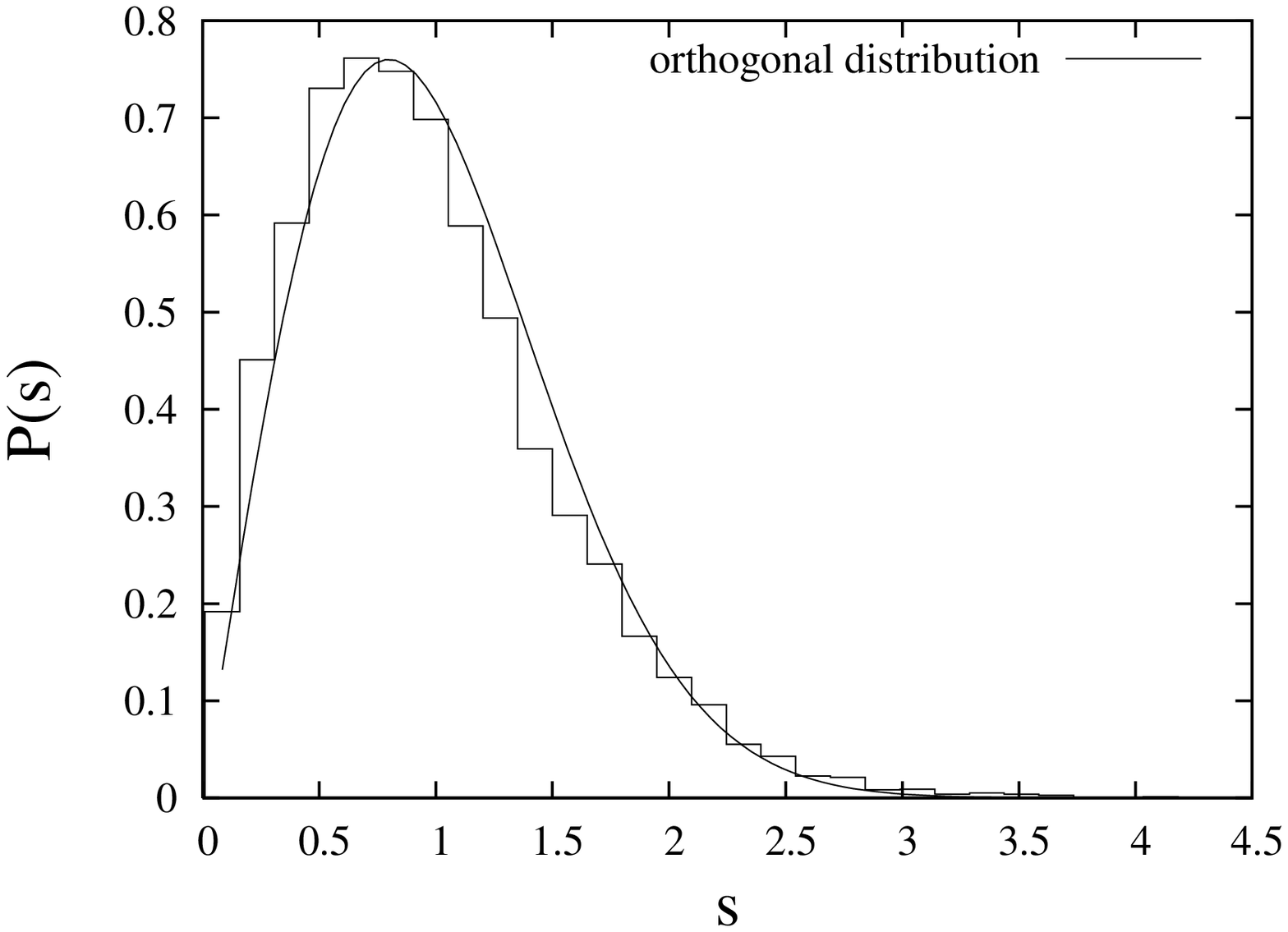}}
\caption{Unfolded eigenvalue spacing distribution for $\beta = 2.35,\,
L=18$.}
\label{spacing1_fig}
\end{figure}

    It is also of interest to study the eigenvalue spacing distribution of the ``unfolded"
spectrum \cite{unfolding},  which has previously been studied, in connection with
random matrix theory, for the Dirac spectrum of SU(2) and SU(3) pure gauge theory. 
The unfolded spectrum is a rescaling of the spectrum such that the average spacing
between neighboring eigenvalues is unity, and a quantity of interest is the distribution
of fluctuations around that average value.  The procedure \cite{su2unfolding} is as follows:
Let $\l_{i}^{n}$ be the $i$-th eigenvalue of the covariant Laplacian in the $n$-th
lattice configuration, with $\l_{i}^{n}>\l_{i-1}^{n}$.  The set of all $\{\l_{i}^{n}\}$ of
all configurations is then sorted in ascending order, and we denote by $N_i^n$ the location (from $1$ to
$N_{conf}\times N_{ev}$) of $\l_{i}^{n}$ in the sorted list.  For the n-th configuration, 
the eigenvalue spacing $s^{n}_{i}$ between neighboring eigenvalues in the unfolded spectrum 
is defined as
\beq
           s^{n}_{i} = {N_{i+1}^{n} - N_{i}^{n} \over N_{conf}}
\eeq 
 Let $P(s)$ denote the probability distribution of these spacings (i.e.\ regardless of
 $i,~n$).  In the case of the Dirac operator \cite{su2unfolding}, the distribution $P(s)$ is well
described by a certain Wigner distribution. For the spectrum of the
covariant Laplacian,  our numerical results show that eigenvalue spacing distribution of the
unfolded spectrum is also well described by one of the Wigner distributions, namely, 
the orthogonal distribution
\begin{equation}
P(s) =  {\pi\over 2} \, s \, {\rm e}^{-{\pi\over 4}s^2}\, , \label{spacing}
\label{Ps}
\end{equation}
as seen in Fig.~\ref{spacing1_fig} for $\b=2.35$ on an $18^{4}$ lattice volume. 
We have also checked that eq.\  \rf{Ps} gives a good fit to the eigenvalue spacing
distribution at $\beta=2.4$ and $\beta=2.3$.

     Returning to the density of states $\r(\l)$ of the original spectrum,
we find that the quadratic ($\a=2$) power behavior continues well past
$\l_{int}$, but does not persist thoughout the spectrum.   Fig.\ \ref{nz} 
showed an excellent cubic fit to
the cumulative $n(z)$ data, but this was for low-lying eigenmodes with $n(z)<12$.  
In Figs.\   \ref{pc25v16}  and \ref{pc25v20} we show all of the data
for $n(z)$, obtained at $\b=2.5$ on $16^{4}$ and $20^{4}$ lattices,
together with the cubic fit to the data at all volumes.  The vertical line 
in each graph indicates the value of $z$ which corresponds, at each volume,
to $\l - \l_{0}=0.05$.    In both cases the data for $n(z)$ begins to diverge away from 
the cubic fit at about this value of $\l-\l_{0}$.  In physical units,   
$\D \l = (\l-\l_{0})/a^{2}$ at $\l - \l_{0}=0.05$ is roughly two and a  half times our estimate for 
$\D \l_{int}$ at this coupling.   

\FIGURE[htb]{
\centerline{\includegraphics[width=8truecm]{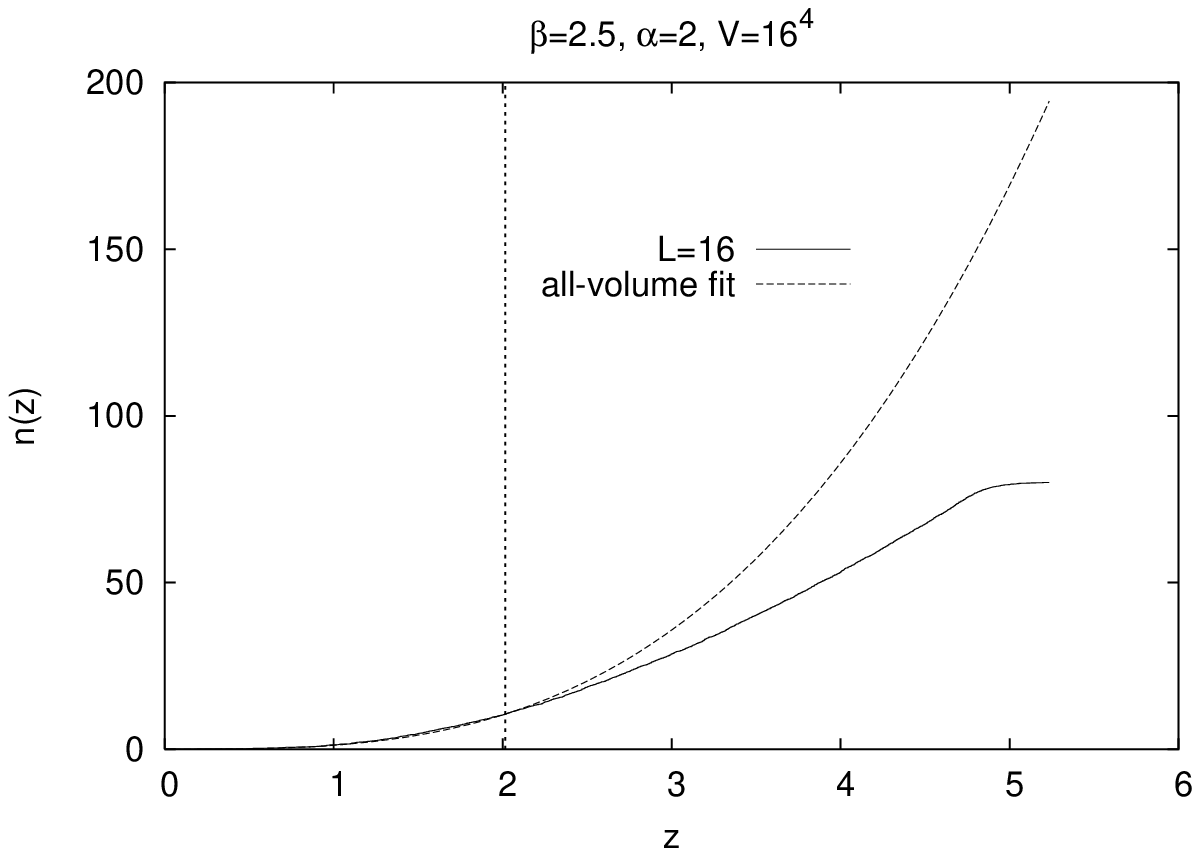}}
\caption{Cumulative eigenvalue number $n(z)$ vs.\ $z$
at $\b=2.5,\a=2$ on a $16^{4}$ lattice volume.  The data shown
includes all calculated eigenmodes, which extend
beyond the mobility edge.  For reference, the vertical line intersects the x-axis
at a value of $z$ corresponding to $\l-\l_{0}=0.05$ (see text).} 
\label{pc25v16}
}

\FIGURE[htb]{
\centerline{\includegraphics[width=8truecm]{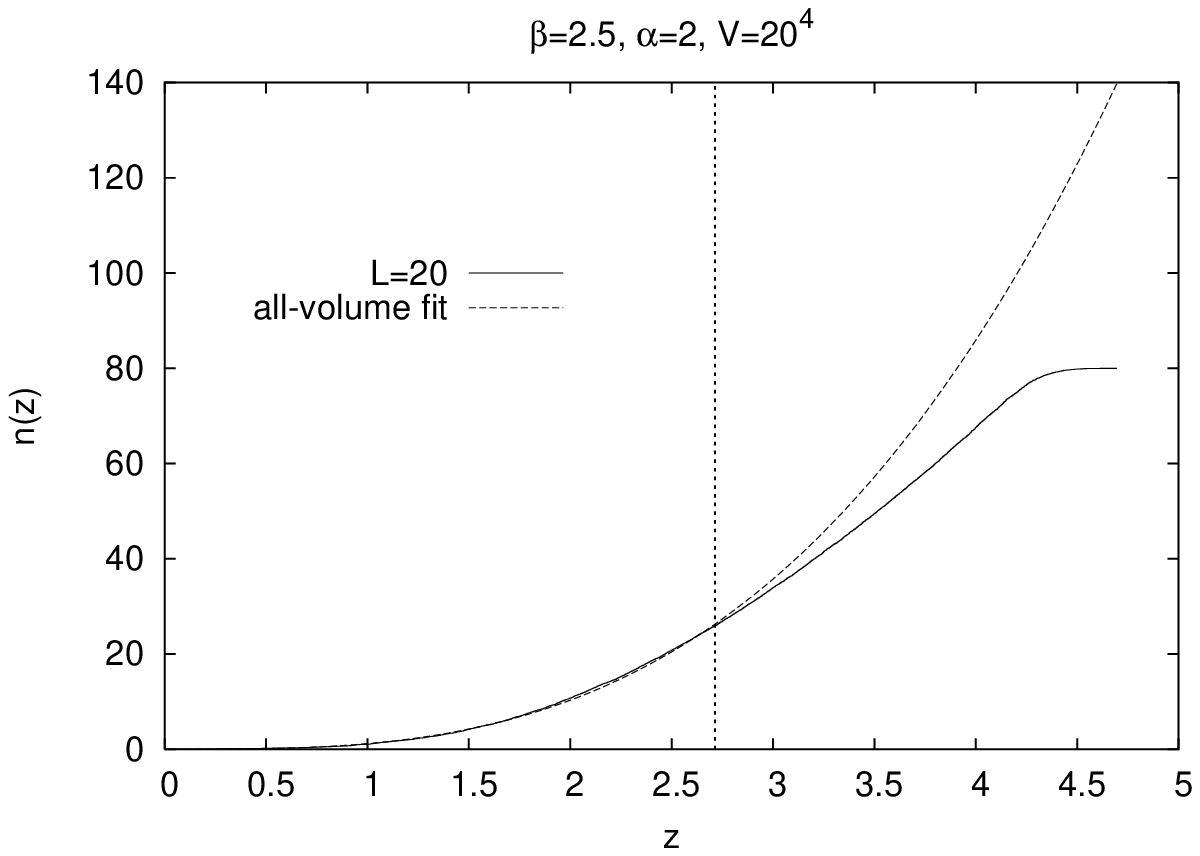}}
\caption{Same as Fig.\ \ref{pc25v16}, on a $20^{4}$ lattice volume.} 
\label{pc25v20}
}

    Given the density of states we can compute the phase-space volume, in physical units, taken 
up by states below the mobility edge.  For purposes of comparison we begin with 
the free case, where it is easy to  show that the density of states in $D=4$ dimensions 
is $\r(\l)= \l/\pi^{2}$.  The total number of
states in the eigenvalue interval  from $\l=0$ to $\l_{stop} = \L a^{2}$ is 
\bea
              N_{\L} &=& \sum_{\l \le \L a^{2}} 1
\non \\
                        &=& V \int_{0}^{\L a^{2}} d\l ~ {\l\over \pi^{2}}
\eea
where $a$ is the lattice spacing, and $\L$ is some fixed number in units of (GeV)${}^{2}$.
Now divide by the physical volume, to get the number of states per unit volume,
$\eta$, which lie in the eigenvalue interval $\l/a^{2} < \L$:
\bea
                 \eta &\equiv& {N_{\L} \over V_{phys}}
\non \\
                   &=& {1\over a^{4}} \int_{0}^{\L a^{2}} d\l ~  {\l\over \pi^{2}}
\non \\
                    &=& {1\over \pi^{2}} \int_{0}^{\L} d\l_{phys} ~ \l_{phys}
\non \\
                    &=&  \oh  {1\over \pi^{2}} \L^{2} 
\eea
Then if $\L$ is a finite cutoff in physical units, $\eta$ is also finite, even as
$a \ra 0$.

    We make the same computation for localized states of the covariant Laplacian,
which lie in the eigenvalue interval $\l\in[\l_{0},\l_{mob}]$ 
\bea
                 \eta &\equiv& {N_{\l_{mob}} \over V_{phys}}
\non \\
                   &=& {1\over a^{4}} \int_{\lm}^{\l_{mob}} d\l ~ \k(a) (\l-\lm)^{2}
\non \\
                    &=&  {\k(a)\over 3} {(\l_{mob}-\lm)^{3} \over a^{4}}
\eea
Denoting eigenvalues in physical units as $\l^{phys}$
\beq
              \l = \l^{phys} a^{2}
\eeq
we have
\beq
           \eta = {\k(a)\over 3} a^{2} (\l^{phys}_{mob}-\lm^{phys})^{3}
\eeq
According to the data in the previous section, 
\beq
           \l^{phys}_{mob}-\lm^{phys} = M/a
\eeq
where $M\approx 0.045$ GeV.
This means that 
\beq
          \eta = {1\over a} {\k(a) \over 3} M^{3}
\eeq

     The question is then how the factor $\k$
in the density of states  \rf{rho} varies as a function of $a$.
In the fit to the $n(z)$ data at $\b=2.5$, we found $\k \approx 3.8$.
Repeating the same analysis at $\b=2.35$, we find $\k \approx 2.8$.
Thus $\k$ does not seem to decrease with smaller lattice spacing; if anything,
it increases.

     From this data, we draw a
remarkable conclusion:   The number of states below the mobility edge, per unit 
physical volume, diverges in the continuum limit.   This means that if we compute 
the contribution to any observable, keeping only a finite 
number of the lowest scalar field eigenmodes per unit volume, then 
\emph{all} of the contributing states would be localized states in the continuum limit.
This is an unexpected feature of adjoint scalar fields in confining gauge theories, and
suggests that the particle propagator is dominated, at scales which are large compared
to the lattice spacing, by the localized eigenmodes.  In the next section we present a 
second argument leading to the same conclusion.

\section{Localization and the Scalar Propagator}\label{discussion}

      In a free theory, the eigenmodes of scalar and fermionic kinetic operators 
are simply plane wave states, and this is also the starting point of the
weak-coupling perturbative approach to gauge theories.  In perturbation theory, 
it is assumed that the eigenmodes of kinetic operators such as the covariant 
Laplacian, in a suitable (e.g.\ Landau) gauge, are approximately plane wave states, 
at least for Euclidean momenta which are large compared to $\Lambda_{QCD}$.
On the other hand, we have seen in section \ref{interval}  that for the
adjoint covariant Laplacian, the interval between the lowest eigenvale $\l_{0}$ and 
the mobility edge $\l_{mob}$ tends to infinity, in physical units, in the continuum
limit.  What this implies is that the contribution of finite momenta extended modes 
to the scalar particle propagator is negligible, unless we are willing to tolerate a mass subtraction
which would introduce tachyonic modes into the theory.

        On the lattice, the scalar particle propagator in the quenched (no scalar loop) 
approximation has the form
\beq
             G^{ab}(x-y) = \sum_{n} \left\langle
                     {\p^{a}_{n}(x) \p^{b}_{n}(y) \over \l_{n} + m_{0}^{2} } \right\rangle
\eeq
where the VEV is evaluated in the pure gauge theory with some appropriate gauge
choice.  By ``finite momentum extended modes", we mean  
extended eigenmodes ($\l>\l_{mob}$) whose Fourier components are negligible for momenta
$|\vec{p}|/a < P$, where $P$ is a momentum cutoff which is large in physical units, 
but small compared to $1/a$. 
Consider the contribution to the scalar propagator, in the continuum limit, due to 
these extended, finite momenta eigenmodes.
Denoting this contribution by $G'(x-y)$, we have in physical units
\beq
             G_{phys}^{'ab}(x-y) = {1\over a^{2}}{\sum_{n}}' \left\langle
                     {\p^{a}_{n}(x) \p^{b}_{n}(y) \over \l_{n} + m_{0}^{2} } \right\rangle
\eeq
where $\sum'_{n}$ denotes the sum over the finite momenta, extended eigenmodes.
The number of such eigenmodes cannot exceed the number of lattice momenta $\vec{p}$
satisfying the restriction  $|\vec{p}|/a < P$.  This number is of order $a^{4} V \D V^{phys}_{P}$,
where $\D V^{phys}_{P}=\oh\pi^{2}P^{4}$ is the momentum space volume in physical  units,
and $V$ is the lattice volume in lattice units. 
We can easily estimate the magnitude of $G'_{phys}$ from
\bea
              {\sum_{n}}' 1 &\sim& a^{4} V \D V^{phys}_{P}
\non \\
               \p^{a}_{n}(x) \p^{b}_{n}(y) &\sim& {1\over V} 
\eea 
so that        
\beq
           G'^{ab}(x-y) \sim {\D V_{p}^{phys} \over (\l' +m_{0}^{2})/a^{2}}
\label{G'}
\eeq               
where $\l'$ is the magnitude of a typical eigenmode in the range of eigenmodes
summed by $\sum'_{n}$.  On the other hand, in the continum limit, $\l'/a^{2} \ra \infty$,
and therefore the contribution to the scalar propagator from finite momentum
extended modes is negligible, for any $m_{0}^{2}>0$.  This is by no means special 
to confining gauge theories; it is also true in QED.  The resolution in QED is to
allow for a negative bare mass term $m_{0}^{2}<0$, adjusted so that
$\l_{0} + m_{0}^{2}$ is $O(a^{2})$.    But for the adjoint scalar in Yang-Mills
theory, this choice of counterterm is inadequate.  The problem is that for
eigenvalues $\l_{n} > \l_{mob}$  contributing to $G'(x-y)$
\beq
           {\l_{n} + m_{0}^{2}\over a^{2}} > {\l_{0} + m_{0}^{2}\over a^{2}} + \D \l_{mob} 
\eeq
Therefore, even if we chose a counterterm such that $\l_{0}^{2} + m_{0}^{2} = 0$,
the denominator in eq.\ \rf{G'} would still be of order $\D\l_{mob} \sim 1/a$, which
means that $G' \ra 0$ in the continuum limit.  The only way to avoid this is to
choose a counterterm such that $\l_{mob}+m_{0}^{2}$ is $O(a^{2})$.  But in that
case, the eigenvalues $\l_{n}+m_{0}^{2}$ of the kinetic operator for the localized 
eigenmodes $\l_{n}<\l_{mob}$ are negative, i.e.\ tachyonic.

    In a quenched theory the bare mass constant can be chosen at will, and
tachyon modes in the scalar kinetic term are not excluded.   However, in a well-defined field 
theory with a dynamical scalar field  there can be no true tachyon modes; these only 
appear in perturbative calculations around a false vacuum state.  Thus the
quenched scalar propagator can only approximate the scalar propagator in the
unquenched theory when the lowest eigenmode $\l_{0} + m_{0}^{2}$ of the kinetic
operator is, on average, greater than zero.
 
     In D=4 dimensions, SU(2) gauge-Higgs theory with the scalar field
in the adjoint representation (aka the Georgi-Glashow model) is known to have two 
distinct phases: a confinement phase and a Higgs phase \cite{adj-higgs}.\footnote{In contrast, 
a gauge-Higgs theory with the scalar in the fundamental representation
has only one phase, and the asymptotic string tension is vanishing for all finite gauge and
Higgs couplings (cf.\ ref.\ \cite{fund-higgs}).}    Our quenched calculation would be relevant as
an approximation to the scalar propagator of SU(2) gauge-Higgs theory in the confined phase.
In this theory
\bea
          Z &=& \int DU \int D\vph ~ e^{-(S_{YM}+S_{\p})}
\non \\
             &=& \int DU ~ e^{-(S_{YM} + \D S)}
\eea
where
\bea
     S_{YM}[U] &=& -\b \sum \oh \mbox{Tr}[UUUU] 
\non \\
       S_{\vph}[\p,U] &=& \sum \Bigl[\oh \vph (-D)\p 
                   + \oh m_{bare}^{2} \vph^{2} + \g \vph^{4}]   
\non \\
        e^{-\D S[U]} &=& Z_{\p}[U] = \int D\vph ~ e^{-S_{\p}[\p,U]}            
\eea
The covariant Laplacian and scalar field are in the color adjoint representation;
all color indices are implicit.   

Define expectation values in a fixed background gauge field as
\beq
          \langle Q \rangle_{U} \equiv {1\over Z_{\vph}} \int D\vph ~ Q[\vph,U] e^{-S_{\vph}}
\eeq
and in particular, 
\beq
    \langle \vph^{a}(x) \vph^{b}(y)\rangle = 
            {\int DU ~ \langle \vph^{a}(x) \vph^{b}(y)\rangle_{U} ~ e^{-(S_{YM} + \D S)} \over
                \int DU ~ e^{-(S_{YM} + \D S)} }
\eeq
In the quenched approximation, it is assumed that $\D S$ is only a small correction to $S_{YM}$
in the confined phase of the theory.

   As before, let $\p_{n}(x)$ denote the eigenstates of the covariant Laplacian for fixed $U$,
and expand the scalar field  
\beq
           \vph^{a}(x) = \sum \P_{n} \p^{a}_{n}(x)
\eeq
Defining
\beq
           {1\over \L_{n}} \equiv  \langle \P_{n} \P_{n}  \rangle_{U}  
\eeq
and assuming 
\beq
           \langle \P_{n} \P_{m}  \rangle_{U} \approx 0 ~~~\mbox{for}~~m\ne n 
\eeq
we have
\beq
         \langle \vph^{a}(x) \vph^{b}(y)\rangle = 
         \sum_{n} \left\langle {\p^{a}_{n}(x) \p^{b}_{n}(y) \over \L_{n}} \right\rangle
\eeq
         
         The quenched approximation is based on the assumption that  
it is possible to choose some values for $\b$ and $m_{0}^{2}$ in the quenched theory, 
dependent on the gauge coupling and the Higgs couplings $m_{bare}^{2},\g$ in the
confined phase of gauge-Higgs theory, such that
\beq
\sum_{n} \left\langle {\p^{a}_{n}(x) \p^{b}_{n}(y) \over \L_{n}} \right\rangle_{gauge-Higgs} \approx
\sum_{n} \left\langle {\p^{a}_{n}(x) \p^{b}_{n}(y) \over \l_{n} + m_{0}^{2}} \right\rangle_{pure~YM}
\label{approx}
\eeq
where the expectation values on the lhs and rhs are taken in the unquenched and pure Yang-Mills
theories respectively.   The argument for the validity of this approximation is the usual one: large-N,
and the general expectation that (i) as long as the system remains in the confined phase, the effect
of matter loops on the vacuum state is not very drastic; and (ii) the main effect of the 
$\g \p^{4}$ term on the scalar particle propagator is to renormalize the mass term.
  
      In gauge-Higgs theory, the $\L_{n}$ must, from their definition, be positive semi-definite; the
quenched approximation (essentially $\langle \L_{n}\rangle \approx \langle \l_{n} + m_{0}^{2}\rangle$) 
can only be relevant to the unquenched theory for $\langle \l_{n}+m_{0}^{2} \rangle \ge 0$.   
In that case, as we have seen, the contribution of finite momentum extended modes to the
scalar propagator is negligible, and it is the localized modes which dominate scalar particle propagation
at distances which are large compared to the lattice scale.  This would mean that ordinary weak-coupling
perturbation theory goes very wrong for adjoint scalar particles, even at distance scales which are quite
small compared to the confinement scale.\footnote{Another short-distance phenomenon which is missed by
perturbation theory is the excess lattice action found on thin P-vortex sheets and monopole lines;
c.f.\  ref.\ \cite{finetune}.}

\section{Conclusions} 

    In a previous article \cite{Us} we had found something odd 
about the spectrum of the covariant Laplacian in the adjoint 
representation:  the lowest eigenmodes appear to be localized 
in a volume which shrinks to zero, in physical units, in the continuum 
limit.   In the present work we have extended our study to the full 
interval of localized states, and have found other surprising 
features:  First, the range of eigenvalues $\D \l_{mob}$ of the localized 
eigenmodes tends to infinity, in physical units, in the continuum limit.    
Secondly, the density of eigenmodes rises from zero quadratically, up to the mobility
edge (suggestive of a connection to multi-critical matrix 
models of degree $m=1$), and the number of localized eigenmodes per unit physical 
volume is infinite in the continuum limit.  We must add a caveat that 
these conclusions rest on the results of numerical simulations of SU(2) 
lattice gauge theory, carried out at couplings between $\b=2.2$ and 
$\b=2.5$, and lattice volumes up to $24^{4}$.  It is, of course, not excluded 
that the trend in the data could change at higher couplings and/or 
larger volumes.

      An infinite range of eigenvalues between the lowest eigenvalue and the mobility edge has 
an interesting consequence.  If mass counterterms leading to tachyonic modes 
are excluded, then the quenched scalar particle propagator is dominated by localized 
modes; there is a negligible contribution from extended eigenmodes corresponding
to finite physical momenta. Exclusion of tachyonic modes 
is necessary if the quenched propagator is to be a reasonable approximation 
to the scalar propagator in a gauge theory with dynamical scalar fields.
Again there is a caveat:  We do not really know if an infinite range of localized 
eigenmodes persists in the unquenched theory.  This will require further, 
computationally more intensive, investigations.

      If the propagator for adjoint scalar fields is completely dominated by localized states, 
when evaluated in the confined phase of gauge-Higgs theory in some appropriate gauge,
then this would raise some doubts about the
validity of perturbation theory, at least as applied to adjoint scalar fields.  
Although one naturally expects weak-coupling perturbation theory to break down at a distance scale
comparable to the confinement scale, it is generally believed that this procedure should
provide correct answers for short-distance quantities.   If, however, the scalar propagator
is dominated, even at short distances, by localized eigenmodes, then weak-coupling
perturbation theory may be misleading.   This is an interesting (and obviously radical) possibility, 
which calls for further investigation.

%
\acknowledgments{%
Our research is supported in part by the U.S. Department of Energy
under Grant No.\ DE-FG03-92ER40711 (J.G.) and
the Slovak Science and Technology Assistance Agency under Contract
No.\ APVT--51--005704 and the Grant Agency for Science, Project VEGA
No.\ 2/6068/2006 (\v{S}.O.). 
The work of the ITEP group (A.V.P., M.I.P. and S.N.S.) was partially supported by
grants RFBR-05-02-16306a, RFBR-0402-16079, DFG-RFBR 436 RUS 113/739/2, RFBR-DFG
06-02-04010 and EU Integrated Infrastructure Initiative Hadron Physics (I3HP)
under contract RII3-CT-2004-506078 (MIP), RFBR-05-02-16306a and
RFBR-05-02-17642 (S.M.M. and A.V.K.). \\ 

\appendix*
\section{Alternative Fits}

    In addition to the linear fit of IPR to $\l$, we have explored some other
possible fits and fitting procedures for estimating $\l_{mob}$.  In particular
we have tried to fit various fractional powers of the IPR to $\l$, and to bin the data in
different ways.   As an example,  we have subdivided
the interval between minimal and maximal value of $\lambda$ into 10
subintervals and calculated the average IPR${}^{1/p}$ of the eigenvalues in each
interval. We associate, with each average IPR${}^{1/p}$, a value of $\l$ in
the center of the corresponding $\l$ interval, and then fit these data points to
\begin{equation}\label{IPRfit}
{\left(\frac{\mbox{IPR}}{L^4}\right)}^{1/p} = A\lambda+B
\end{equation}
The x-axis intercept $\l_{int}$ 
\begin{equation}
\lambda_{int} = -\frac BA\, .
\end{equation}
is taken to be an estimate of the mobility edge $\l_{mob}\approx \l_{int}$.

\FIGURE[t!]{
\centerline{\includegraphics[width=8truecm]{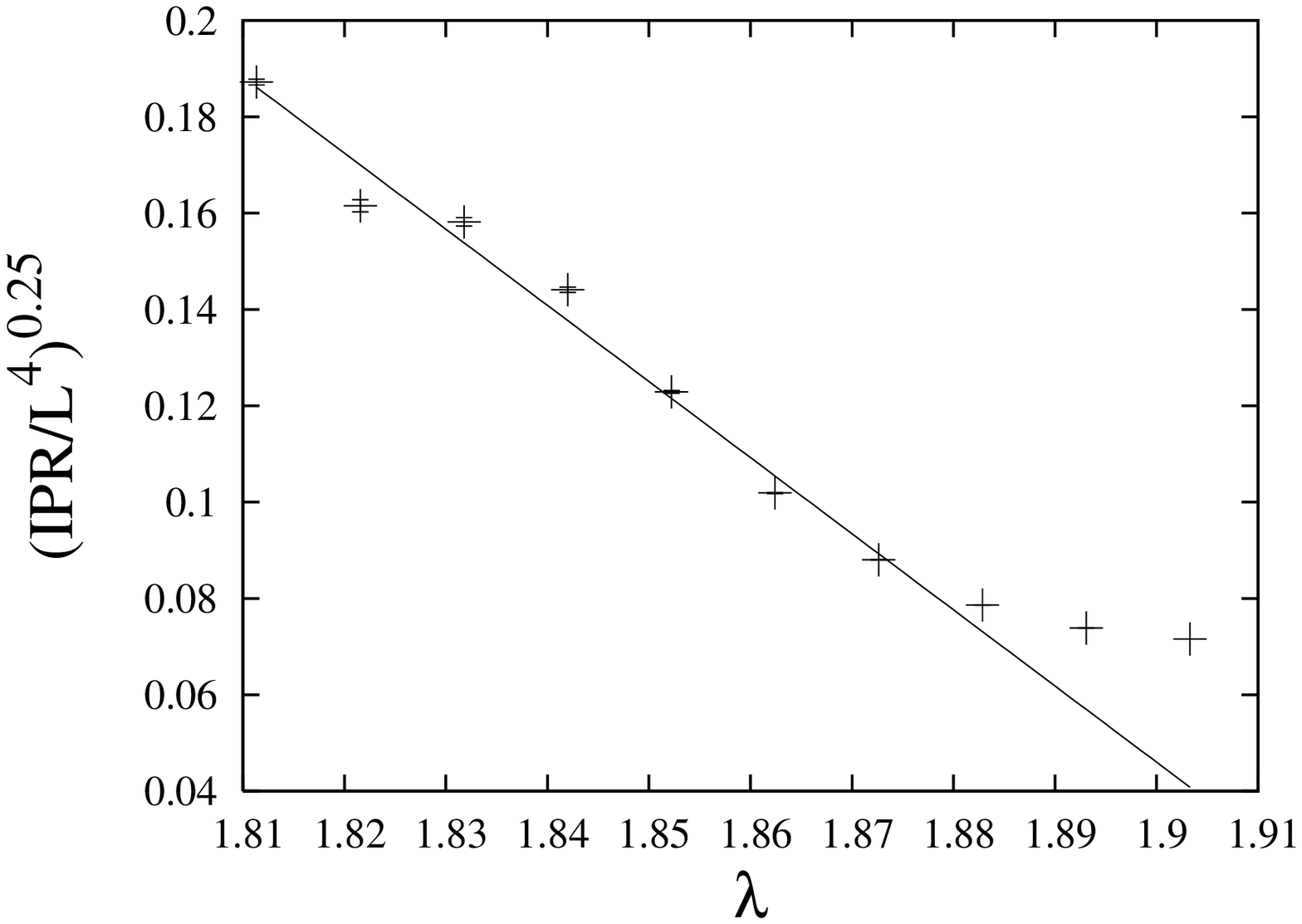}}
\caption{Dependence of ${\left(\frac{\mbox{IPR}}{L^4}\right)}^{0.25}$ on
$\lambda$ for $\beta = 2.40,\, L=20$.} 
\label{IPRfig}
}

\FIGURE[t!]{
\centerline{\includegraphics[width=8truecm]{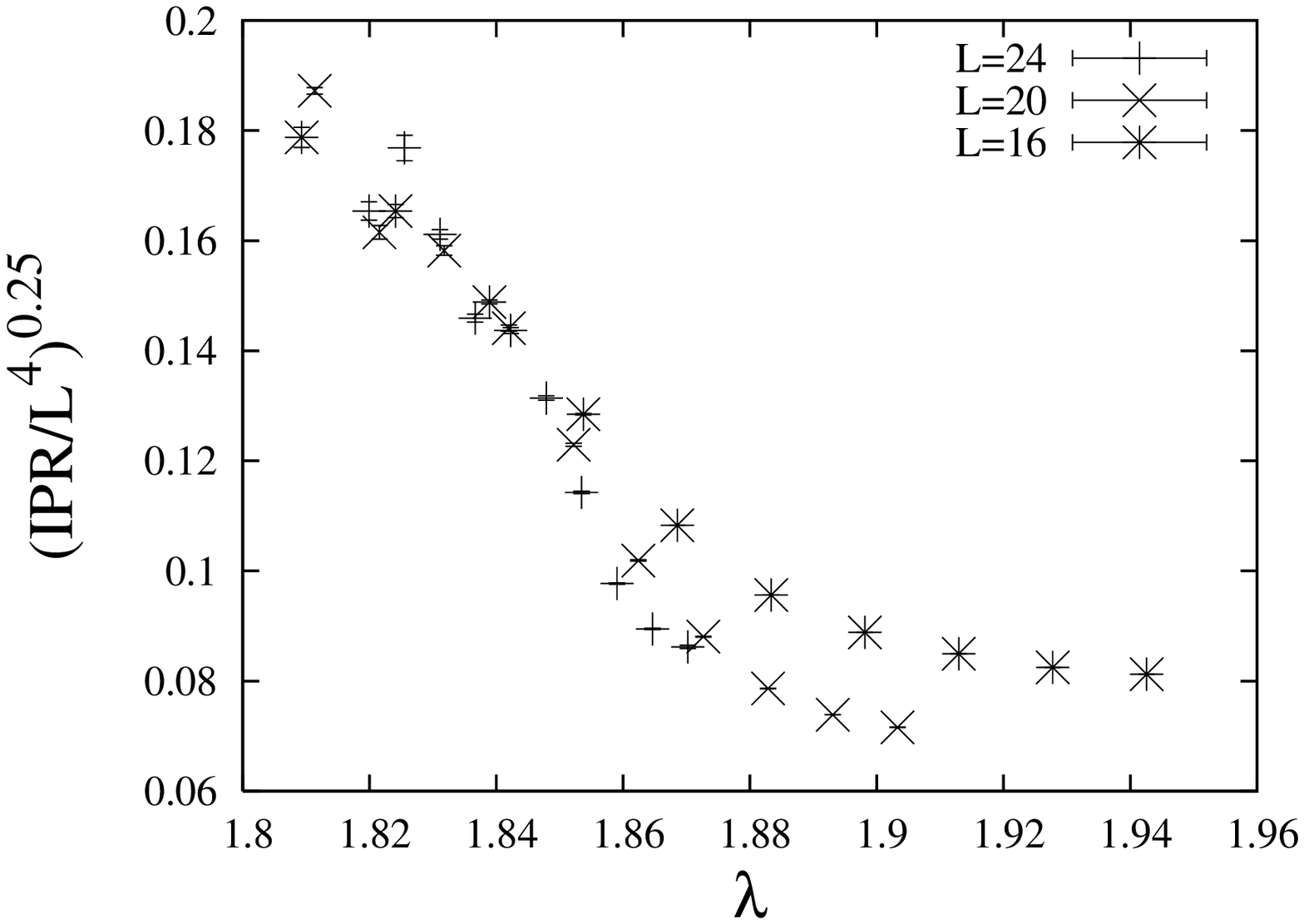}}
\caption{Dependence of ${\left(\frac{\mbox{IPR}}{L^4}\right)}^{0.25}$ on
$\lambda$ for various $L$ and $\beta = 2.40$.} 
\label{volume}
}

    A plot of $(IPR/L^{4})^{1/4}$ vs.\ $\l$ at $\b=2.4$, on a $20^{4}$ lattice is shown in Fig.\ \ref{IPRfig};
the data obtained at $\b=2.4$ on $16^{4},~20^{4},~24^{4}$ lattice volumes are displayed together in
Fig.\ \ref{volume}.   We should note that these plots tend to underestimate $\l_{0}$, in that the range
of $\l$ (which is divided into ten intervals) depends on the highest and lowest values of $\l_{n}$ found in all
lattice configurations.   This means that lattice configurations in which the lowest eigenvalues are well below 
the average may introduce data points in the graph whose lowest $\l$ value
is also well below the average $\l_{0}$.    Nevertheless, a linear fit of $IPR^{1/p}$ vs.\ $\l$, with $p\ne 1$, 
might conceivably be superior to a linear fit of $IPR$ vs.\ $\l$ in the neighborhood of the mobility edge,
and this in turn would give a better estimate for $\l_{mob}$.
We have concluded, however, after some experimentation with different fits,  that the data is not really 
adequate to convincingly determine the optimal value of $p$ near the mobility edge.

%
%

\end{document}